\documentclass[12pt]{article}
\usepackage{amsmath,amssymb,amsfonts,mathrsfs}
\usepackage[pdftex]{graphicx}
\usepackage{amsmath,amssymb,amsfonts}
\usepackage[pdftex]{graphicx}
\usepackage{bm}
\usepackage{dcolumn}
\def\d{\partial}
\def\half{\frac{1}{2}}
\def\D{\mathscr{D}}
\def\Z{\mathcal{Z}}
\def\beq{\begin{equation}}
\def\eeq{\end{equation}}
\def\bea{\begin{eqnarray}}
\def\eea{\end{eqnarray}}
\def\A{\vec{A}(x)}
\def\ru{\hat r}
\def\cbf{\vec{c}}
\title{\textbf{Two-potential theory of electric and magnetic charges via duality transformation}}
\author{Chandrasekhar Chatterjee$^{a}$\footnote{E-mail: chandra@imsc.res.in},
Indrajit Mitra$^{b}$\footnote{E-mail: indrajit.mitra@saha.ac.in, imphys@caluniv.ac.in}~ and
H. S. Sharatchandra$^{a}$\footnote{E-mail: sharat@imsc.res.in} \\\\
$^a$ The Institute of Mathematical Sciences, C.I.T. Campus, Taramani P.O.,\\
Chennai 600113, India\\
$^b$ Department of Physics, University of Calcutta,\\
92 A.P.C. Road, Kolkata 700009, India}
\date{}
\begin{document}
\maketitle
\begin{abstract}
Dirac, Schwinger and Zwanziger theories of electric and magnetic charges are obtained via duality transformation. 
Analogous construction for three Euclidean dimensions, with magnetic charges interacting with electric currents, is also done.
The role of Dirac strings as dislocations in the configurations of gauge potential is emphasized.
\end{abstract}


\section{\bf Introduction}
In this paper we obtain Dirac, Schwinger and Zwanziger theories \cite{dirac, Zwanziger:1968rs} of electric and magnetic 
charges via duality transformation. Our aim is to build  unified techniques for handling
a field and its dual on an equal footing. The reason is that the dual field plays an important role 
in many contexts. It behaves as a disorder parameter and drives the properties of the theory in some phases. 
The dual field couples locally to certain topological configurations of the original field. Therefore by 
keeping them we will be able to naturally handle some non-perturbative aspects  of the theory.
A plasma or condensate of such topological configurations may be qualitatively and quantitatively 
relevant in some phases. By having a formalism which has both fields, we can handle such effects on the original field.
Also combinations of the field and its dual close together often have exotic properties and 
play  crucial role for the properties of the theory.

  We illustrated our techniques for a scalar field in two Euclidean dimensions in \cite{Chatterjee:2011vd}. The end point of a line discontinuity
is the source for the dual field, much like the end point of a Dirac string behaving as a magnetic monopole in electrodynamics \cite{dirac}. 
The two-dimensional local Lagrangian involving both the scalar field and its dual  \cite{kane}  is the analogue of Zwanziger's \cite{Zwanziger:1968rs}
two-potential local theory of electric and magnetic charges \cite{cabibbo,singleton}. 
The dislocation line becomes invisible for a quantization of dual charges as in Dirac theory of magnetic monopoles.
The correlation of the field with its dual has unusual properties.

   All these show that the two-potential formalism  is not restricted to electrodynamics.
Any field theory can be recast as a local theory of the fields and their duals present together. This is important for 
quantum chromodynamics because confinement property is expected to be driven by topological configurations such as monopoles
 and vortices. We need to be able to see their effects on gluons and quarks. Therefore a formalism with both fields together
 is very useful.
 
 The scheme of this paper is as follows. We use Euclidean formalism throughout this paper to highlight the role played 
 by  $ \sqrt{-1} $. In Sec.\ \ref{sec:2}, we start with a real massless scalar field in three Euclidean dimensions. We relate it to
 Abelian gauge theory by a duality transformation. Point sources for the scalar field are mapped  to Dirac strings acting as
dislocation lines in the configurations of the gauge potential. We further relate this to a local theory with both 
the scalar and gauge potential present simultaneously and coupling locally to the magnetic charges and electric currents.
We demonstrate how the interactions amongst magnetic charges and electric currents is recovered. In Sec.\ \ref{sec:3},
we begin with the Abelian gauge theory in three Euclidean dimensions and recover the same formalism. Here the electric 
currents act as sources of surface dislocations in the configuration of magnetic scalar potential.
In Sec.\ \ref{sec:4}, we obtain the Dirac, Schwinger and Zwanziger formulations of electric and magnetic charges via duality
transformation in four Euclidean dimensions. We discuss the relevance of our techniques in Sec.\ \ref{sec:5}.
In the Appendix, we give some techniques useful for handling the Dirac potential of a magnetic monopole.

\section{\label{sec:2}Line dislocation and a local action with both the `photon' and the `dual photon' in three Euclidean dimensions} 

We begin with a free massless real scalar field $\chi(x)$ in three Euclidean dimensions. 
Its correlation functions can be obtained from the functional integral
\begin{eqnarray}
\Z[\rho] = N_1 \int\D\chi(x)\, e^{\int d^3x\left[-\frac{1}{2} 
\left(\vec\nabla\chi(x)\right)^2 + i \rho(x) \chi(x)\right]}\,. 
\label{N1} 
\end{eqnarray}
Here $ N_1 $ is a normalization factor such that $ \Z[\rho=0] = 1 $, and  
 $ \rho(x) $ is an external source coupling locally to $ \chi(x) $. We have 
 deliberately included $ \sqrt{-1} $ in this source term for convenience in performing the duality
 transformation below. We linearize the dependence on $ \chi(x) $ in the exponent in equation (\ref{N1}) 
 by introducing an auxiliary field $ \vec{B}(x) $:
 \begin{eqnarray}
\Z[\rho] = N_2 \int \D\vec{B}(x)\D\chi(x) e^{\int d^3x\left[ -\half \vec{B}(x)^2 - i \vec{B}(x)\cdot\vec\nabla\chi(x) +
i\rho(x)\chi(x)\right]}\,.
\label{N2}
\end{eqnarray}
A formal integration over $ \chi(x) $ gives
\begin{eqnarray}
\Z[\rho] = N_3 \int \D\vec{B}\prod_{\vec{x}}\delta(\vec\nabla\cdot\vec{B}(x) + \rho(x))e^{\int d^3x\left[ -\half \vec{B}(x)^2\right]}\,.
\label{N22}
\end{eqnarray}
For a point source 
\begin{eqnarray}
\rho(x) = g \delta^3(\vec{x} - \vec{y}),
\label{ps1}
\end{eqnarray}
the $ \delta $-functional constraint in (\ref{N22}) corresponds to a magnetic monopole of strength $ g $ at the point $ \vec{y} $. 
For solving it, we choose the particular integral in the form of a Dirac string from $ \vec{y} $ along the $-$ve $ z $-direction. This 
singular solution gives a net flux $ g $ through any surface enclosing $ \vec{y} $. Thus the solution for a general $\rho(x)$ is 
\begin{eqnarray}
\label{monopole1}
\vec{B}(x)= \vec\nabla\times\vec{A}(x) - \hat{n_3} \d_3^{-1}\rho(x)
\end{eqnarray}
where
\begin{eqnarray}
 \d_3^{-1}\rho(x)= - \int_{x_3}^\infty\, dx'_3 \,\rho(x_1, x_2, x'_3)
 \label{monopole2}
\end{eqnarray}
and $\hat{n_3}$ is the unit vector in the 3-direction. 
The use of equation (\ref{monopole1}) in equation (\ref{N22}) rewrites the massless scalar theory of (\ref{N1}) as 
an Abelian gauge theory. The gauge field has one transverse degree of freedom in 
three Euclidean dimensions, matching that of the scalar theory.

We shall refer to $\vec{A}(x)$ as the `photon' and $\chi(x)$  as the `dual photon'.  We are interested in their mutual
correlations. We therefore  include a source $ \vec{j}(x) $ for $ \vec{A}(x) $:
\begin{eqnarray}
\label{N4}
\Z[\rho,\vec{j}] = N_4\int\D\vec{A}\,e^{\int d^3x\left[-\half\left(\vec\nabla\times\vec{A}(x) - \hat{n_3} \d_3^{-1}\rho(x)\right)^2
+ i\vec{j}(x)\cdot\vec{A}(x)\right]}.
\end{eqnarray}
The dual photon $ \chi(x)$ couples locally to the magnetic monopole density $\rho(x)$. Thus equation (\ref{N4}) gives a (gauge) 
theory with both electric current and magnetic charges. Equation (\ref{N4}) shows that a point magnetic charge at $ \vec{y} $
 has the effect of a line dislocation (the Dirac string) starting at $ \vec{y}$ in the configuration space of 
the gauge potential.

 As a consequence of these singular dislocation lines, the configurations $ \vec{A}(x) $ which matter in the functional
 integral (\ref{N4}) are not the usual plane waves. For the action to be finite, $ \vec\nabla\times\vec{A} $ 
  should also be singular and cancel the Dirac string singularities. Thus the configurations which matter are precisely 
the Dirac potential $ \vec{A}^D(\vec{x}-\vec{y})$ due to a magnetic monopole at $ \vec{y} $ and its distortions.
This is explicitly seen as follows.
Let us shift $ \vec{A}(x) $ to $\vec{a}(x)$ as
\begin{eqnarray}
\vec{A}(x) &=&   \vec{a}(x) +  \int d^3y \vec{A}^D(\vec{x}-\vec{y})\rho(y) \,.
\label{Aa}
\end{eqnarray}
Now use the representation of $ \vec{A}^D(\vec{x}-\vec{y})$ in the form of Green
function for $\partial_3 \nabla^2$, as given in (\ref{Axx'}). This leads to
\begin{eqnarray}
\vec\nabla\times\vec{A}(x) &=& \vec\nabla\times\vec{a}(x) + \frac{1}{4\pi} \int d^3y \frac{\vec{x}-\vec{y}}{|\vec{x}-\vec{y}|^3}\rho(y) + \hat{n_3}
\d_3^{-1}\rho(x).
\label{AaB}
\end{eqnarray}
So this  shift cancels the Dirac string in (\ref{N4}). The second term on the r.h.s of (\ref{AaB}) is simply the 
magnetic field at $ \vec{x} $ due to a magnetic monopole density $ \rho(y) $. It is the gradient $ -\vec\nabla\chi(x) $ of a scalar potential
\begin{eqnarray}
\chi(x) &=&  \int d^3y\, \Delta(\vec{x} - \vec{y}) \rho(y),\\
\Delta(\vec{x} - \vec{y}) &=& \frac{1}{4\pi}\frac{1}{|\vec{x}-\vec{y}|}.
\label{delta}
\end{eqnarray}
Therefore
\begin{eqnarray}
\Z[\rho, \vec{j}] = N_4\int\D\vec{a}\, e^{\int d^3x\left[-\half\left(\vec\nabla\times\vec{a}(x) -\vec\nabla\chi(x)\right)^2
+ i\vec{j}(x)\cdot\vec{a}(x)\right] + i\int d^3x d^3y \vec{j}(x)\cdot\vec{A}^D(\vec{x}-\vec{y})\rho(y)}\,.
\label{aphi1}
\end{eqnarray}
 Now
\begin{eqnarray}
\left(\vec\nabla\times\vec{a}(x)-\vec\nabla\chi(x)\right)^2 = (\vec\nabla\times\vec{a}(x))^2
-2\vec\nabla\cdot(\chi(x) \vec\nabla\times \vec a(x))
+\vec\nabla\cdot(\chi\vec\nabla\chi)+\chi(x)\rho(x)
\label{aphi2}
\end{eqnarray}
as $\nabla^2\chi(x)=-\rho(x)$.
There is no boundary contribution from the total divergence terms. So we get
\begin{eqnarray}
\label{N4a}
\Z[\rho, \vec{j}] = N_4\int\D\vec{a} e^{\left[\int d^3x\left(-\half\left(\vec\nabla\times\vec{a}(x)\right)^2
+ i\vec{j}(x)\cdot\vec{a}(x)\right) -\int d^3x d^3y\left(\half\rho(x)\Delta(\vec{x} - \vec{y})\rho(y) 
- i \vec{j}(x)\cdot\vec{A}^D(\vec{x}-\vec{y})\rho(y)\right)\right]}
\end{eqnarray}
which has a conventional action for the new fluctuations $\vec{a}(x)$. This completes our contention 
that the configurations that contribute to (\ref{N4})
are the Dirac potential $  \int d^3y \vec{A}^D(\vec{x}-\vec{y})\rho(y) $ and its fluctuations. We may 
fix a gauge for $ \vec{a}(x) $ as usual
and integrate over $ \vec{a}(x) $ in (\ref{N4a}). This gives
\begin{eqnarray}
\label{rj}
\Z[\rho, \vec{j}] = N_5 \, e^{\int d^3x d^3y\left[-\half j^i(x)\Delta_{ij}(\vec{x}-\vec{y})j^j(y)
-\half\rho(x)\Delta(\vec{x}-\vec{y})\rho(y) 
+ i \vec{j}(x)\cdot\vec{A}^D(\vec{x}-\vec{y})\rho(y)\right]}\,,                            \label{monocurr}
\end{eqnarray}
where $ \Delta(\vec{x}-\vec{y}) $ and $ \Delta_{ij}(\vec{x}-\vec{y}) $ are respectively the propagator for a massless 
real scalar and the Abelian gauge potential  in three Euclidean 
dimensions \cite{bmk}. 
It shows the electric currents $ \vec{j}(x) $ interacting via the Biot-Savart law.
(Gauge fixing permits us to extend the law to currents that need not be conserved.) It also has the magnetic monopoles 
interacting via the Coulomb potential. In addition it shows that the magnetic monopoles $ \rho(x) $  interact with electric current 
$ \vec{j}(x) $ through the Dirac potential $ \vec{A}^D(\vec{x}-\vec{y}) $. ( Note the $ \sqrt{-1} $
 in this term in equation (\ref{rj}), even in our Euclidean theory. It is not strange as the 
 interaction of the current $ \vec{j} $(x) with a gauge  potential $ \vec{A}(x) $ is given by $ i\vec{j}(x)\cdot\vec{A}(x)$ even in the
 Euclidean theory.) This represents the net effect  of the line discontinuity (the Dirac string) in the 
 configurations of the gauge potential. Only the end point of the string matters and behaves like a magnetic monopole.
 
     The magnetic charge density $ \rho(x) $ has a non-local coupling to the `photon' field $ \A $ in  (\ref{N4}), 
though it couples locally to the dual photon $ \chi(x) $ in (\ref{N1}). We now present a local action that has both the 
photon and the dual photon fields present together. For this we rewrite (\ref{N4}) introducing an auxiliary field $\vec{b}(x)$:
\begin{eqnarray}
\label{bA}
\Z[\rho, \vec{j}] = N_6\int\D\vec{b}\D\vec{A}\, e^{\int d^3x\left[ -\half \vec{b}(x)^2 + i
\vec{b}(x)\cdot \left(\vec\nabla\times\vec{A}(x) - \hat{n_3} \d_3^{-1}\rho(x)\right) +i \vec{j}(x)\cdot\vec{A}(x)\right]}\,.
\end{eqnarray}
Thus $ \rho(x) $ couples locally to $ \d_3^{-1}b_3(x) $, which is to be identified with the dual photon $ \chi(x) $:
\begin{eqnarray}
\chi(x) = \d_3^{-1}b_3(x)\,.
\end{eqnarray}
We may integrate back over $ b_1(x) $ and $ b_2(x) $, to get
\begin{eqnarray}
\Z[\rho, \vec{j}] = N_7\int\D\chi\D\vec{A}\, e^{\int d^3x\left[ -\half (\d_3\chi(x))^2 - \half
 \left(\hat n_3\times(\vec\nabla\times\vec{A}(x))\right)^2 + \,i\d_3\chi(x)\hat n_3\cdot \vec\nabla\times\vec{A}(x) + 
 i \vec{j}(x)\cdot\vec{A}(x) + i\rho(x)\chi(x)\right]}\,.
\label{phiA1}
\end{eqnarray} 

This gives  the local field theory of electric currents and magnetic charges in
three Euclidean dimensions. It is the analogue of the
two-potential formalism in four dimensions \cite{Zwanziger:1968rs} and of the local field theory involving
the scalar field and its dual in two dimensions \cite{Chatterjee:2011vd,kane}. 
Note the following unusual features: 
\begin{itemize}
\item The action is not manifestly rotation invariant. Nevertheless, the rotation covariance 
is restored for physical observables when the Dirac quantization condition for electric and magnetic charges is met.
(See below.)
\item The `kinetic energy' terms for $ \chi(x) $ and $ \A $ have derivatives only in some directions. However,
if we integrate over $ \chi(x) $ (correspondingly $ \A $), we recover the conventional action for $ \A $(correspondingly $ \chi(x) $).
\item The action in (\ref{phiA1}) is not real (even with the imaginary sources switched off). The term bilinear in $ \chi$ and $ \A $ 
is pure imaginary. 
\end{itemize}

The `propagators' can be calculated using Fourier modes, as follows.
In the action of equation (\ref{phiA1}), we use the identity
$(\hat{n_3}\times (\vec\nabla\times\vec A))^2=(\vec\nabla\times\vec A)^2
       -(\hat{n_3}\cdot (\vec\nabla\times\vec A))^2$. 
The resulting $-\frac{1}{2}(\vec\nabla\times\vec A)^2$  term in the action becomes
$\frac{1}{2}\vec A\cdot\nabla^2 \vec A$, upon adding  
the gauge-fixing term corresponding to the Feynman gauge. 
Also $\hat{n_3}\cdot (\vec\nabla\times\vec A)=
(\hat{n_3}\times\vec\nabla)\cdot\vec A$, which is $i\vec k_\perp\cdot\vec A(k)$
in momentum space,
where $\vec k_\perp\equiv \hat{n_3}\times\vec k$. Thus we obtain
\bea
\Z [\rho, \vec j]&=& N_7\int \D\chi(k)\D\vec A(k) 
\exp \int d^3k \Big(-\frac{1}{2}\chi(-k){k_3}^2\chi(k)
-\frac{1}{2} A_i(-k)(k^2 \delta_{ij}-{k_\perp}_i{k_\perp}_j)A_j(k)          \nonumber\\
&&+iA_i(-k){k_\perp}_i k_3 \chi(k)
+i j_i(-k)A_i(k)+i \rho(-k) \chi (k)
\Big)\,.
\eea
The propagators are then obtained by the inversion of a matrix:
\begin{eqnarray}
\label{U}
\left[\begin{tabular}{ll}
                                              \phantom{xx} $k_3^2$ & \phantom{xxx}$ -ik_3 k_{\perp i}$ \\
                                               $-ik_3 k_{\perp j}$  & $ 
                                   ~~k^2 \delta_{ij} - k_{\perp i}k_{\perp j}$\\ 
                         \end{tabular}\right]^{-1} =  \frac{1}{k^2} \left[ \begin{tabular}{ll} 
                                                                            $ 1 $   & $ i\frac{k_{\perp i}}{k_3} $ \\
                                            $ i\frac{k_{\perp j}}{k_3} $  & $ \delta_{ij}$\\                
                                                                           \end{tabular}\right] .
\end{eqnarray}
In position space, the propagators are
\begin{eqnarray}
\langle \chi(x)\chi(y)\rangle &=& \Delta(\vec{x}-\vec{y})\\
\langle A_i(x)A_j(y)\rangle &=& \Delta_{ij}(\vec{x}-\vec{y}) = \delta_{ij}\Delta(\vec{x}-\vec{y}) \\
\label{Aphicorrelation}
\langle A_i(x)\chi(y)\rangle &=& iA^D_i(\vec{x}-\vec{y})
\end{eqnarray} 
where $\Delta(\vec{x}-\vec{y})$ is given by (\ref{delta}), and we have used (\ref{Amom}).
Alternatively, we can read off these position space propagators from (\ref{rj}).

Even though $ \chi(x) $ and $ \A $ are real fields, the propagator $\langle A_i(x)\chi(y)\rangle $ is pure imaginary.
This is possible because the action is not real. We see that $ iA^D_i(\vec{x}-\vec{y}) $ serves as the `propagator'
connecting the electric currents and magnetic charges. The correlation of $ \chi(x) $ with the `magnetic field' $ \vec{B}(x) = \vec\nabla\times\A $
has the Dirac string singularity:
\begin{eqnarray}
\langle \vec{B}(x)\chi(0)\rangle = i \left(\frac{1}{4\pi}\frac{\vec x}{|\vec x|^3} + \hat{n}_3 \delta(x_1)\delta(x_2)\theta(-x_3)\right)\,.
\end{eqnarray}
Because of the explicit presence of the Dirac string, rotation invariance in (\ref{Aphicorrelation}) and (\ref{monocurr})
 is not manifest. Dirac\cite{dirac} argued that with a `quantization'  of electric ($ e $) and magnetic ($ g $) charges, 
 the Dirac string becomes invisible and rotation covariance is restored. Consider a point magnetic charge given by 
(\ref{ps1}), and a loop $ C $ carrying a current
 \begin{eqnarray}
j_i(x) = e\oint_C d\tau \frac{dX_i(\tau)}{d\tau}\delta^3(\vec{x} - \vec X(\tau))         \label{j_i(x)}
\end{eqnarray} 
where $\tau$  is an arbitrary parametrization of the loop $ C $. 
The contribution to the cross-correlation of $ \exp\left[ie\oint dx^iA_i(x)\right]$ with $\exp\left[ig\chi(y)\right]$
comes from the last term in the exponent of (\ref{monocurr}) for the sources (\ref{ps1}) and (\ref{j_i(x)}).
(This is the analogue of the cross-correlation between the vertex operators for the scalar field and the dual field
in two dimensions, which was considered in \cite{Chatterjee:2011vd}.) Thus this cross-correlation
equals $ \exp\left[ieg\oint dx^iA^D_i(\vec x-\vec y)\right]$. Using Stokes' theorem, this is
 \begin{eqnarray}
\exp\left(-i\frac{eg}{4\pi}\Omega(C)\right)
\end{eqnarray}
where $ \Omega(C) $ is the solid angle subtended by $ C $ at the site $\vec y$ of the magnetic charge, and the solid angle
is to be computed by  using a surface bounding $ C $ which does not intersect the Dirac string. 
Therefore when an infinitesimal loop $C$ does not enclose the string, we get the contribution 1, but 
when it encloses the string, 
we get the contribution $e^{-ieg}$. Only with the quantization condition
\bea
eg = 2\pi n                \label{eg2pi}
\eea
the latter contribution is also 1, and the Dirac string is then invisible to any  current loop.
\section{\label{sec:3}Surface dislocations and scalar potential theory of electric currents and magnetic charges
in three Euclidean dimensions}

In Sec.\ \ref{sec:2}, we started with a massless real scalar field and obtained the two-potential theory of 
magnetic charges interacting with electric currents in three Euclidean dimensions. In this Section, 
we begin with Abelian gauge theory and  obtain the same two-potential formalism. This exercise is 
instructive for the case of four Euclidean dimensions. 

We begin with
\begin{eqnarray}
\label{ZA3}
\Z[\vec{j}] = N_8\int\D\A e^{\int d^3x\left[-\half (\vec\nabla\times\A)^2 +\, i \vec{j}(x)\cdot\A\right]}
\end{eqnarray}
describing the current $\vec{j}(x)$ interacting via the gauge potential $ \A $ in three Euclidean dimensions. 
Rewriting 
\begin{eqnarray}
\label{bA3}
\Z[\vec{j}] &=& N_9\int\D\vec{b}\D\vec{A}\, e^{\int d^3x\left[ -\half \vec{b}(x)^2 + i
\vec{b}(x)\cdot \vec\nabla\times\vec{A}(x) + i \vec{j}(x)\cdot\vec{A}(x)\right]}\\
&=& N_{10} \int \D\vec{b}\prod_{\vec{x}}\delta(\vec\nabla\times\vec{b}(x) +
\vec{j}(x))e^{\int d^3x\left[ -\half \vec{b}(x)^2\right]}\,.           \label{bbb}
\end{eqnarray} 
The consistency of the constraint requires 
\begin{eqnarray}
\label{conservation}
 \vec\nabla\cdot\vec{j}(x) = 0. 
\end{eqnarray}
Choosing the solution
\begin{eqnarray}
b_i(x) =  \d_i\chi(x) - \epsilon_{3il}\d_3^{-1}j_l(x)
\label{bi}
\end{eqnarray}   
for the $ \delta $-functional constraint, we get
\begin{eqnarray}
\label{chij}
\Z[\rho, \vec{j}] = N_{11}\int \D\chi\, e^{\int d^3x 
\left[-\half(\d_1\chi(x)- \d_3^{-1}j_2(x))^2 
-\half (\d_2\chi(x)+ \d_3^{-1}j_1(x) )^2 -\half(\d_3\chi(x))^2 + i\rho(x)\chi(x)\right]}
\end{eqnarray}
where we now introduced the source for $ \chi $. 
Note that the component $ j_3(x) $ is not explicitly present. 
However, by the conservation law (\ref{conservation}), we can write
\begin{eqnarray}
\label{j3}
j_3(x) = - \d_3^{-1}(\d_1j_1(x) + \d_2j_2(x))   \label{con2}
\end{eqnarray}
and therefore it is implicitly present. 

Equation (\ref{chij}) is giving the interaction of magnetic charges
and electric currents using the scalar potential $ \chi(x) $ encountered in magnetostatics.
Consider a current loop $ C $ in the 1-2 plane with a charge $e$ flowing in it. Equation (\ref{chij}) 
shows that for the action to be finite in this case, there should be a discontinuity 
in the scalar potential $ \chi $ in the form of a surface dislocation. This dislocation is along 
a cylindrical domain wall with $ C $ as the mouth and extending all the way to infinity in the 3-direction.
The gradient of the potential jumps by $ e $ across the domain wall. This is the conventional description
of using a multivalued magnetic scalar potential in the presence of electric currents \cite{kleinert}.

We linearize (\ref{chij}) in a specific way:
 \begin{eqnarray}
\Z[\rho, \vec{j}] =N_{12} \int \D\chi \D\A \prod_x \delta(A_3(x) - \alpha_3(x)) \exp\int_x\Big[
-\half(\d_3\chi(x))^2 - \half(\d_2 A_3(x) - \d_3 A_2(x))^2   \nonumber\\
  - \half(\d_3 A_1(x) - \d_1 A_3(x))^2 -i (\d_2 A_3(x) - \d_3 A_2(x))(\d_1\chi(x) - 
 \d_3^{-1} j_2(x)) \nonumber\\
- i (\d_3 A_1(x)  - \d_1 A_3(x))(\d_2\chi(x) + \d_3^{-1}j_1(x)) + i\rho(x)\chi(x)\Big]
\label{Achij}
\end{eqnarray} 
Equation (\ref{chij}) is reproduced from equation (\ref{Achij}) by shifting $A_I$, $I=1,2$ to 
$ A'_I = A_I - \d_3^{-1} \d_IA_3$ and integrating over $A'_I$.  The potential $A_3$ has been
introduced in this step to have a gauge-invariant Lagrangian. The gauge-fixing condition $ A_3(x) = \alpha_3(x)$
corresponding to the axial gauge is here the simplest choice as it is not affected by the shift from $A_I$ to $A'_I$.

Finally it can be checked, using (\ref{con2}), that the action in (\ref{Achij}) is equal to the action in (\ref{phiA1}).
The Lagrangian being gauge-invariant, we can now pass from the axial gauge to any other gauge we find convenient.

Instead of choosing an infinite domain wall, we can  simply choose a finite surface $ S $
enclosing the current loop $C$ to be the dislocation for the scalar potential. For that case, in the place of (\ref{bi}),
we have
\begin{eqnarray}
b_i(x) = \d_i\chi(x)+ \d_i\chi_s(x),
\label{bi2}
\end{eqnarray} 
where $ \chi_s $ in discontinuous across the surface $ S $:
\begin{eqnarray}
\vec\nabla{\chi_s}(x) = \int_S d^2X(s)\, \hat{n}(s)\,e \,\delta^3(\vec{x}-\vec{X}(s)),
\end{eqnarray}
$ \vec X(s) $ being a point on the surface $ S $, and $ \hat n(s) $ the normal to $ S $ at this point.
Because of this discontinuity, $(\partial_i\partial_j-\partial_j \partial_i){\chi_s}(x)\neq 0$,
and (\ref{bi2}) gives $\vec\nabla\times\vec b(x)+\vec j(x)=0$ with $\vec j(x)$ given
by (\ref{j_i(x)}). (This can be checked by using the identity
$\int d\vec S\times \vec\nabla\psi=\oint d\vec l\,\psi$ where $\psi$ is a scalar function.)
When (\ref{bi2}) is used in (\ref{bbb}), the finiteness of the action requires $\chi$ to have
a discontinuity across $S$ so as to cancel the discontinuity in $\chi_s$.
Therefore for a loop $ C' $ linked to $ C $, we get 
\begin{eqnarray}
e^{ig\oint_{C'}dx^i\d_i\chi(x)} = e^{iegN_{CC'}}
\label{linkingno}
\end{eqnarray}
where $ N_{CC'} $ is the linking number: the number of times loop $ C' $ winds around
loop $ C $  in the clockwise sense. (This is because the  integral of the current $\vec j$ over any
open surface bounding the loop $C'$ equals $e N_{CC'}$.)
Thus, (\ref{linkingno}) equals 1
when the quantization condition (\ref{eg2pi}) is satisfied.
\section{\label{sec:4}Two-potential theory of electric and magnetic charges in four Euclidean dimensions}
Consider the quantized Abelian gauge field in four Euclidean dimensions:
\begin{eqnarray}
\Z[j_{\mu}] = N_{13}\int \D A_{\mu} e^{\int d^4x\left[-\frac{1}{4}(\d_\mu A_\nu(x) - \d_\nu A_\mu(x))^2 + ij_\mu A_\mu(x) \right]}\,.
\end{eqnarray}
Here $ j_\mu$  ($\mu= 1,2,3,4 $) is the external current. We have 
\begin{eqnarray}
\Z[j_{\mu}] && =  N_{14} \int \D b_i\D e_i \D A_{\mu} \exp \int d^4x\Big[-\half e_i(x)^2  - \half b_i(x)^2 + i e_i(x)(\d_4 A_i(x) 
- \d_i A_4(x))\nonumber\\ && \phantom{xxxxxxxxxxxxxxxxxxxx}
+ i \epsilon_{ijk} b_i(x)\d_j A_k(x) + ij_i(x) A_i(x) + i j_4(x)A_4(x)\Big]\\
&& = N_{15}\int \D b_i\D e_i \prod_x \delta(\d_ie_i(x) + j_4(x))\prod_x\delta(\epsilon_{ijk}\d_j b_k(x)  - \d_4 e_i(x) + j_i(x) ) 
\nonumber\\ &&\phantom{xxxxxxxxxxxxxxxxxxxxxxxxxxxxx} \times\exp \int d^4x\Big[-\half e_i(x)^2  - \half b_i(x)^2\Big]\,.
\label{constraint1}
\end{eqnarray}
From the $ \delta $-functional constraints we have the consistency condition
\begin{eqnarray}
\d_ij_i(x) + \d_4j_4(x) = 0\,.                 \label{con4}
\end{eqnarray}
We solve the first constraint in (\ref{constraint1}) as
\begin{eqnarray}
e_i(x) = \epsilon_{ijk}\d_jC_k(x) - \delta_{i3} \d_3^{-1}j_4(x)\,,
\end{eqnarray}
corresponding to choosing the Dirac string along the 3-direction
(similar to (\ref{monopole1})).
Using this, the second constraint becomes
\begin{eqnarray}
\epsilon_{ijk}\d_j(b_k - \d_4 C_k) = -j_i - \delta_{i3} \d_3^{-1}\d_4j_4\,.
\end{eqnarray}
We solve this in the form (similar to (\ref{bi}))
\begin{eqnarray}
b_k(x) - \d_4C_k(x) = -\d_kC_4(x) - \epsilon_{3kl}\d_3^{-1}j_l(x).
\end{eqnarray}
Therefore we get 
\begin{eqnarray}
\label{C}
\Z[j_\mu,k_\mu] &=& N_{16} \int \D C_\mu \exp\int  d^4x \Big[-\half (\d_1 C_2(x) - \d_2 C_1(x) - \d_3^{-1}j_4(x))^2\nonumber\\  
&&- \half (\d_I C_3(x) - \d_3 C_I(x))^2 
  -\half (\d_I C_4(x) - \d_4 C_I(x) + \epsilon_{IJ}\d_3^{-1} j_J(x))^2\nonumber\\
&&-\half (\d_3 C_4(x) - \d_4 C_3(x))^2 + i k_\mu(x) C_\mu(x) \Big]
\end{eqnarray}
where we have introduced a source $ k_\mu $ for $ C_\mu $. Here the indices $ I,J $ run over only $1$ and $2$.
This gives the Dirac and Schwinger formulations. As in Sec.\ \ref{sec:4}, we linearize (\ref{C}) in a particular way.
The first term in the Lagrangian is linearized to
\bea
-\half (\d_3 A_4(x) - \d_4 A_3(x))^2+i(\d_3 A_4(x) - \d_4 A_3(x))
(\d_1 C_2(x) - \d_2 C_1(x) - \d_3^{-1}j_4(x))\nonumber
\eea
while the third term is linearized to
\bea
-\half(\d_3 A_I(x) - \d_I A_3(x))^2 +i\epsilon_{IJ}(\d_3 A_I(x) - \d_I A_3(x))
(\d_J C_4(x) - \d_4 C_J(x) + \epsilon_{JK}\d_3^{-1} j_K(x))\,.\nonumber
\eea
The axial gauge condition is imposed on $A_3$.
(We get back (\ref{C}) by shifting $A_4$ and $A_I$, and then integrating.)
This linearized form of (\ref{C}),
on using (\ref{con4}), is equal to
\begin{eqnarray}
\Z[j_\mu,k_\mu] &=& N_{17}\int \D C_\mu \D A_\mu  \prod_x \delta(A_3(x) - \alpha_3(x)) \exp\int d^4x \Big[-\half (\d_I A_3(x) - \d_3 A_I(x))^2  \nonumber\\ 
&& -\half (\d_3 A_4(x) - \d_4 A_3(x))^2 - \half (\d_I C_3(x) - \d_3 C_I(x))^2 
 -\half (\d_3 C_4(x) - \d_4 C_3(x))^2 \nonumber\\ && 
+ i (\d_3 A_4(x)-\d_4 A_3(x))(\d_1 C_2(x) - \d_2 C_1(x))\nonumber\\
&& +  i \epsilon_{IJ} (\d_3 A_I(x) - \d_I A_3(x))(\d_J C_4(x) - \d_4 C_J(x))
+ ij_\mu(x) A_\mu(x) + i k_\mu(x) C_\mu(x) \Big]\,.\nonumber\\
 \label{dual4D}
\end{eqnarray}
Thus we have recovered Zwanziger's two-potential theory of electric and magnetic changes via a duality transformation. 
If we had chosen $ \hat{n} $ as the direction of the Dirac string instead of the 
$-$ve z-axis, we would have 
got for the exponent on the r.h.s of (\ref{dual4D})
\begin{eqnarray}
\int d^4x & \Big[-\half(\hat{n}\cdot\vec{E}(x))^2 -\half(\hat{n}\times\vec{B}(x))^2 - \half(\hat{n}\cdot\vec{\mathcal E}(x))^2 - \half(\hat{n}\times\vec{\mathcal B}(x))^2
 + i(\hat{n}\cdot\vec{\mathcal B}(x))(\hat{n}\cdot\vec{E}(x)) \nonumber\\
& + i(\hat{n}\times\vec{\mathcal E}(x))\cdot(\hat{n}\times \vec{B}(x))
 + ij_\mu(x) A_\mu(x)  + i k_\mu(x) C_\mu(x)\Big],
\end{eqnarray}
where $\vec{B} = \vec\nabla\times\vec A$, 
$~\vec{E} = -\vec\nabla A_4 + \d_4 \vec A$,
$~\vec{\mathcal B} = \vec\nabla\times\vec{C}$ and 
$\vec{\mathcal E}= -\vec\nabla C_4 + \d_4 \vec{C}$.
\section{\label{sec:5}Discussion}
For a variety of reasons, it is useful to have a formulation with both a field and its dual field simultaneously present in a local theory.
In \cite{Chatterjee:2011vd} this was done for a scalar theory in two Euclidean dimensions and the advantages were highlighted. In this paper, we have carried this out for Abelian
gauge theory in three and four Euclidean dimensions.

The general features are:
\begin{itemize}
\item The sources for a field are certain types of singular dislocations in the configurations of the dual field and also vice versa.
\item The role of these dislocations is to force discontinuous boundary conditions on the fields. Thereby new  sectors
of the field configurations are explored.
\item A local theory with both the field and its dual present simultaneously has certain unusual
features. Though there are more fields, it is equivalent to the original theory and the degrees of freedom are 
not changed. This happens because the dual fields are hidden in the auxiliary fields as specific non-local combinations.
As a consequence correlations of fields with their duals have unusual properties. The theory is not manifestly rotation invariant. However,
rotation covariance is recovered for the `right'   observables with a quantization of the charges of the field and the dual field.
  \item  These features are already known in the context of Dirac's theory of magnetic monopoles. 
Our thrust is that they are general properties of dual fields and not restricted to electrodynamics. We have obtained Dirac, Schwinger and Zwanziger
formulations of electric and magnetic charges via duality transformations.
 We have emphasised the role of Dirac string as singular dislocation in the configurations of the
  gauge  potentials. These issues are relevant for non-Abelian gauge theory. Many non-perturbative aspects such as confinement
  are expected to be driven by topological configurations which couple locally to the dual field. This will be discussed elsewhere. 
\end{itemize}

\section*{Acknowledgement}

I.M. thanks UGC (DRS) for support.
 
\appendix
\leftline{\null\hrulefill\null}\nopagebreak
\section{\label{appendix}Appendix}
In this Appendix, we represent the Dirac vector potential of a monopole in the form of a Green function.
The Dirac potential of a monopole located at the origin has the form
\begin{eqnarray}
\vec A^D(x)= \frac{1 }{4\pi}\frac{\sin\theta}{r(1+\cos\theta)}\hat{\phi} = \frac{1 }{4\pi}\hat{n}_3\times\frac{\ru}{r + x_3} 
\label{Dp}
\end{eqnarray}
with the Dirac string along the $-$ve $z$-direction.
Here $r=|\vec x|$ and $\hat r=\vec x/|\vec x|$..
[For checking (\ref{Dp}) and other
results below, a useful formula is
$\hat{n}_3=\cos\theta\ru-\sin\theta\hat{\theta}$.]
Let us write
\begin{eqnarray}
\vec{A}^D(x) = \hat{n}_3\times \cbf                                                \label{Ac}
\end{eqnarray}
where the vector field $\cbf$ is undetermined upto
addition of a vector in the 3-direction. We choose
\begin{eqnarray}
\cbf=\frac{1 }{4\pi}\frac{\ru+\hat{n}_3}{r+x_3}
\end{eqnarray}
so that \cite{baal}
\begin{eqnarray}
\cbf = \vec\nabla f, ~~~f= \frac{1}{4\pi}\ln(r+x_3)\,.                                       \label{Ag}
\end{eqnarray}
Note that $\d_3 f=1/4\pi r$, and so $\d_3 \nabla^2 f=-\delta(x)$.
Thus the Dirac potential at $x$ due to a monopole at $x'$ can be expressed in terms of
the Green function for the operator $\partial_3 \nabla^2$:
\begin{eqnarray}
\vec{A}^D(x-x')&=&\frac{1}{4\pi}\hat{n}_3\times\vec\nabla\ln(|x-x'|+x_3-x'_3)            \\
        &=&- \hat{n}_3\times\vec\nabla [(\partial_3 \nabla^2) ^{-1}(x-x')]\,.    \label{Axx'}
\end{eqnarray}

[An alternative form of the Dirac potential is $\vec{A}^D(x) = -\hat{\phi}(1/4\pi r)\cot\theta
=-\hat{n}_3\times\ru (x_3/4\pi\rho^2)$, with the Dirac strings along the $\pm z$ directions.
Here $\rho^2={x_1}^2+{x_2}^2$.
This alternative form is half of the sum of the two forms of $\vec{A}^D(x)$,
one having the Dirac string in the $-$ve $z$-direction and the other having
the Dirac string in the $+$ve $z$-direction. 
In this case, we choose $\cbf=(r\hat{n}_3-x_3\ru)/4\pi\rho^2$ in Eq.\ (\ref{Ac}).
Then
$\cbf = \vec\nabla f$ and  $\partial_3 f=1/4\pi r$ continue to hold, but  
with $f=(1/8\pi)\ln((r+x_3)/(r-x_3))$. So equation (\ref{Axx'}) is still valid]

The result given in Eq.\ (\ref{Axx'}) can
also be seen by going over to the Fourier space. For the potential of Eq.\ (\ref{Dp}),
\begin{eqnarray}
\vec\nabla \times \vec{A}^D(x) = \frac{\ru}{4\pi r^2}+\hat{n}_3 \delta(x_1) \delta(x_2)\theta(-x_3)\,.
\end{eqnarray}
Taking the Fourier transform, we get
\begin{eqnarray}
 \vec{k} \times \vec{A^D}(\vec{k})=-\frac{\vec{k}}{k^2}+\frac{\hat{n}_3}{k_3 }\,.
\end{eqnarray}
(The Fourier transform of the theta function can be obtained using
$d\theta(x)/dx=\delta(x)$.) We now evaluate $\vec{k} \times $ both sides and use
$\vec{k}\cdot \vec{A^D}(\vec k)=0$ (since $\vec\nabla\cdot\vec{A}^D(x)=0$) to obtain
\begin{eqnarray}
\vec{A^D}(\vec{k})=  \frac{ \hat{n}_3\times \vec{k}}{k_3 k^2}\,.                  \label{Amom}
\end{eqnarray}
This agrees with Eq.\ (\ref{Axx'}).

\end{document}